\documentclass[preprint2]{aastex}

\usepackage{aastexug}
\usepackage{epsfig}

\begin{document}

\title{Star Formation in the ISO\footnote{Based on observations with ISO, an
ESA project with  instruments funded by ESA Member States (especially the PI
countries:  France, Germany, the Netherlands and the United Kingdom) and
with the  participation of ISAS and NASA.} Atlas of Bright Spiral Galaxies}

\author{George J. Bendo,\altaffilmark{2,3,4}
Robert D. Joseph,\altaffilmark{2,3}
Martyn Wells,\altaffilmark{5} Pascal Gallais,\altaffilmark{6}
Martin Haas,\altaffilmark{7} Ana M. Heras,\altaffilmark{8, 9}
Ulrich Klaas,\altaffilmark{7} Ren\'{e} J. Laureijs,\altaffilmark{8, 9}
Kieron Leech,\altaffilmark{9, 10} Dietrich Lemke,\altaffilmark{7}
Leo Metcalfe,\altaffilmark{9}
Michael Rowan-Robinson,\altaffilmark{11} Bernhard Schulz,\altaffilmark{9, 12}
and Charles Telesco\altaffilmark{13}}

\altaffiltext{2}{University of Hawaii, Institute for Astronomy,
2680 Woodlawn Drive, Honolulu, HI 96822, USA;
bendo@ifa.hawaii.edu, joseph@ifa.hawaii.edu}
\altaffiltext{3}{Visiting Astronomer at the Infrared Telescope Facility,
which is operated by the University of Hawaii under contract from the
National Aeronautics and Space Administration.}
\altaffiltext{4}{Visiting Astronomer at the UH 2.2 m Telescope at Mauna Kea
Observatory, Institute for Astronomy, University of Hawaii.}
\altaffiltext{5}{UK Astronomy Technology Center, Royal Observatory Edinburgh,
Blackford Hill, Edinburgh EH9 3HJ, Scotland, UK; mw@roe.ac.uk}
\altaffiltext{6}{CEA/DSM/DAPNIA Service d'Astrophysique,
F-91191 Gif-sur-Yvette, France; gallais@discovery.saclay.cea.fr}
\altaffiltext{7}{Max-Planck-Institut f\"{u}r Astronomie,
K\"{o}nigstuhl 17, D-69117, Heidelberg, Germany;
haas@mpia-hd.mpg.de, klaas@mpia-hd.mpg.de, lemke@mpia-hd.mpg.de}
\altaffiltext{8}{Astrophysics Division, Space Science Department of ESA,
ESTEC, P.O. Box 299, 2200 AG Noordwijk, Netherlands;
aheras@estsa2.estec.esa.nl, rlaureij@rssd.esa.int}
\altaffiltext{9}{ISO Data Center, Astrophysics Division, ESA,
Villafranca del Castillo, 28080 Madrid, Spain;
lmetcalf@iso.vilspa.esa.es}
\altaffiltext{10}{Said Business School, Park End Street, Oxford OX1 1HP,
England, UK; kieron.leech@said-business-school.oxford.ac.uk}
\altaffiltext{11}{Astrophysics Group, Imperial College, Blackett Laboratory,
Prince Consort Road, London SW7 2BZ, England, UK;
m.rrobinson@ic.ac.uk}
\altaffiltext{12}{IPAC, California Institute of Technology 100-22, Pasadena,
CA 91125 USA; bschulz@ipac.caltech.edu}
\altaffiltext{13}{Department of Astronomy, University of Florida,
P.O. Box 112055, Gainesville, Florida 32611, USA;
telesco@astro.ufl.edu}

\shorttitle{Star Formation in Spiral Galaxies}
\shortauthors{Bendo et al.}

\begin{abstract}
We investigate star formation along the Hubble sequence using the ISO
Atlas of Spiral Galaxies.  Using mid-infrared and far-infrared flux densities
normalized by K-band flux densities as indicators of recent star formation,
we find several trends.  First, star formation activity is stronger in
late-type (Sc~- Scd) spirals than in early-type (Sa~- Sab) spirals.  This
trend is seen both in nuclear and disk activity. These results confirm
several previous optical studies of star formation along the Hubble sequence
but conflict with the conclusions of most of the previous studies using IRAS
data, and we discuss why this might be so.   Second, star formation is
significantly more extended in later-type spirals than in early-type
spirals.  We suggest that these trends in star formation are a result of
differences in the gas content and its distribution along the Hubble
sequence, and it is these differences that promote star formation in
late-type spiral galaxies.  We also search for trends in nuclear star
formation related to the presence of a bar or nuclear activity.  The nuclear
star formation activity is not significantly different between barred and
unbarred galaxies.  We do find that star formation activity appears to be
inhibited in LINERs and transition objects compared to H~{\small II}
galaxies.  The mean star formation rate in the sample is
1.4~M$_\sun$~yr$^{-1}$ based on global far-infrared fluxes.  Combining these
data with CO data gives a mean gas consumption time of
6.4~$\times$~10$^{8}$~yr, which is $\sim$ 5 times lower than the values
found in other studies.  Finally, we find excellent support for the Schmidt
Law in the correlation between molecular gas masses and recent star
formation in this sample of spiral galaxies.
\end{abstract}


\section{Introduction}

\subsection{Scientific Background}

Spiral galaxies have been recognized as sites of on-going star formation
since the definition of the Hubble sequence \citep{h26}.  However the
relation between star formation activity and the location on the Hubble
sequence has been a continuing controversy for the past twenty years.

Early optical studies comparing star formation activity in spiral galaxies
included \citet{s82}, \citet{kk83}, \citet{k83}.  They found a
general trend in star formation activity with Hubble type,
with stronger star formation in Sc spirals than in Sa spirals.
However, \citet{setal83}, using ground-based 10~$\mu$m observations,
demonstrated that star formation activity was not correlated with any
global galaxy property, including Hubble type.  This laid the groundwork
for a debate on trends in star formation along the Hubble
sequence that continues to the present.

With the advent of IRAS, several new studies on this theme were published,
including \citet{detal84}, \citet{dbs87}, \citet{pr89}, \citet{dy91},
\citet{if92}, \citet{tts96}, and \citet{dh97}.  Most IRAS studies found no
trend in star formation along the Hubble sequence, with the exceptions being
\citet{dbs87} and \citet{pr89}, who found that Sa galaxies are generally
weaker sites of star formation than Sc galaxies, and \citet{dh97}, who
found a population of Sa galaxies with relatively enhanced star formation
activity.  But optical and ultraviolet studies on the problem also
continued, with \citet{detal87}, \citet{detal94}, and \citet{ktc94}
all finding trends in increasing star formation along the Hubble
sequence from Sa to Sc galaxies.  To confuse matters further, \citet{yetal96}
found trends in H$\alpha$ surface brightness along the Hubble sequence but
they argued that the trend did not necessarily reflect a true trend in
star formation activity.  Infrared spectroscopy of the C~{\small II} 158~$\mu$m
line performed with ISO by \citet{lvhetal99} also found a trend in increasing
star formation from early- to late-type spiral galaxies, but this survey
had a small sample of only 19 spiral galaxies.

Clearly, opinion is sharply divided on star formation activity along the
Hubble sequence.  Some of the reasons for these discrepant results may
include the following.  1) Extinction is always a concern in optical or
ultraviolet studies, since star formation regions are always dusty.  2) The
IRAS sweep-scan technique is not ideal for measuring the flux from faint,
extended sources.  For faint, extended galaxies the IRAS point source
algorithm may mistake the faint outer regions of the galaxy itself for the
background and therefore undermeasure the flux.  3) To normalize for galaxy
mass, most IRAS studies divide the far-infrared fluxes by B-band fluxes.
Since the B-band fluxes represent the fluxes from the youngest, bluest stars
in any galaxy, this tends to cancel out any trend in star formation activity
in the far-infrared data.  4) The morphological classifications used are
sometimes biased towards star formation.  In particular, the Revised Shapley
Ames (RSA) Catalog \citep{st87} uses star formation activity
in its classification scheme \citep{sb94}, so results on star formation
trends found by any paper using the RSA classifications may merely reflect
the classification criteria.

\subsection{The Design of this Study}

As a fresh approach to this question, we have completed a new survey of
spiral galaxies with the Infrared Space Observatory (ISO) \citep{ketal96} at
mid-infrared and far-infrared wavelengths, supplemented with  ground-based
JHK photometry. The data are presented in the ISO Atlas of Spiral
Galaxies \citet{betal02}  (henceforth
referred to as Paper 1).  This study incorporates several features that
address the concerns mentioned above.  1) We minimize uncertainties related to
extinction by working strictly with infrared data.  2) The ISO data has
several advantages over IRAS data.  ISO has improved sensitivity and angular
resolution that enables us to  distinguish nuclear from disk emission.
Furthermore, ISO operates in a point and integrate mode, and the background
is measured separately from locations pre-selected from IRAS cirrus maps,
so the flux from faint extended galaxies can be measured more accurately
(although the flux measurements are still limited by background structure).
3) We normalize mid-infrared and far-infrared fluxes with K-band fluxes,
which represent more accurately the total stellar population than B-band
fluxes.  4) We use the Third Reference Catalogue of Bright Galaxies (RC3;
\citet{detal91}) for galaxy classification, since the RC3 classifications
are strictly dependent on morphological features such as the bulge to disk
ratio and the tightness of the spiral arms \citep{d59}.

We first summarize the sample selection, observations, and data
reduction, the details of which can be found in Paper 1.
Next, we discuss our diagnostics for studying
star formation activity.  Then we present our results on trends in star
formation along the Hubble sequence, and we compare these results to
previous results and discuss the underlying mechanisms for producing the
trends.  Next, we examine the galaxies for any possible trends in nuclear
star formation related to the presence of a bar or nuclear activity.  We
conclude by calculating quantitative star formation rates and gas
consunsumption times and by comparing these results to those in the
literature.

\section{Data}

\subsection{The Sample}

The galaxies in this sample are a subset of a complete, magnitude-limited set
of galaxies selected from the RSA Catalog.  The sample comprised galaxies
with Hubble types between S0 and Sd and with magnitudes B$_T$~=~12 or
brighter; galaxies in the Virgo Cluster were excluded.  A randomly selected,
subset of these galaxies was observed by ISO based on target
visibility.  This produced a total of 77 galaxies that are representative
of the range of Hubble types in the RSA Catalog.  Additional information on
the sample and its properties is presented in Paper 1.

\subsection{Observations and Data Analysis}

Detailed information on the observations, the data processing, and the
measurements themselves are  presented in Paper 1. In short, the 60, 100, and
180~$\mu$m photometry was obtained with ISOPHOT \citep{letal96} and
processed with PIA 8.0 \citep{getal97}, with additional processing to correct
for PSF effects. The 12~$\mu$m images were taken with ISOCAM \citep{cetal96}
using the LW10 filter, which covered the range 8-15~$\mu$m and were
processed with CIA 3.0 \citep{oetal97}, followed by full processing with CIR,
an implementation of the algorithms of \citet{setal99} realized by P.
Chanial.  The K-band photometric images were mostly taken with QUIRC at the
UH  2.2~m telescope, with additional data from NSFCAM and SPEX at the  NASA
IRTF.

\subsection{Star Formation Diagnostics}

\begin{figure}
\psfig{file=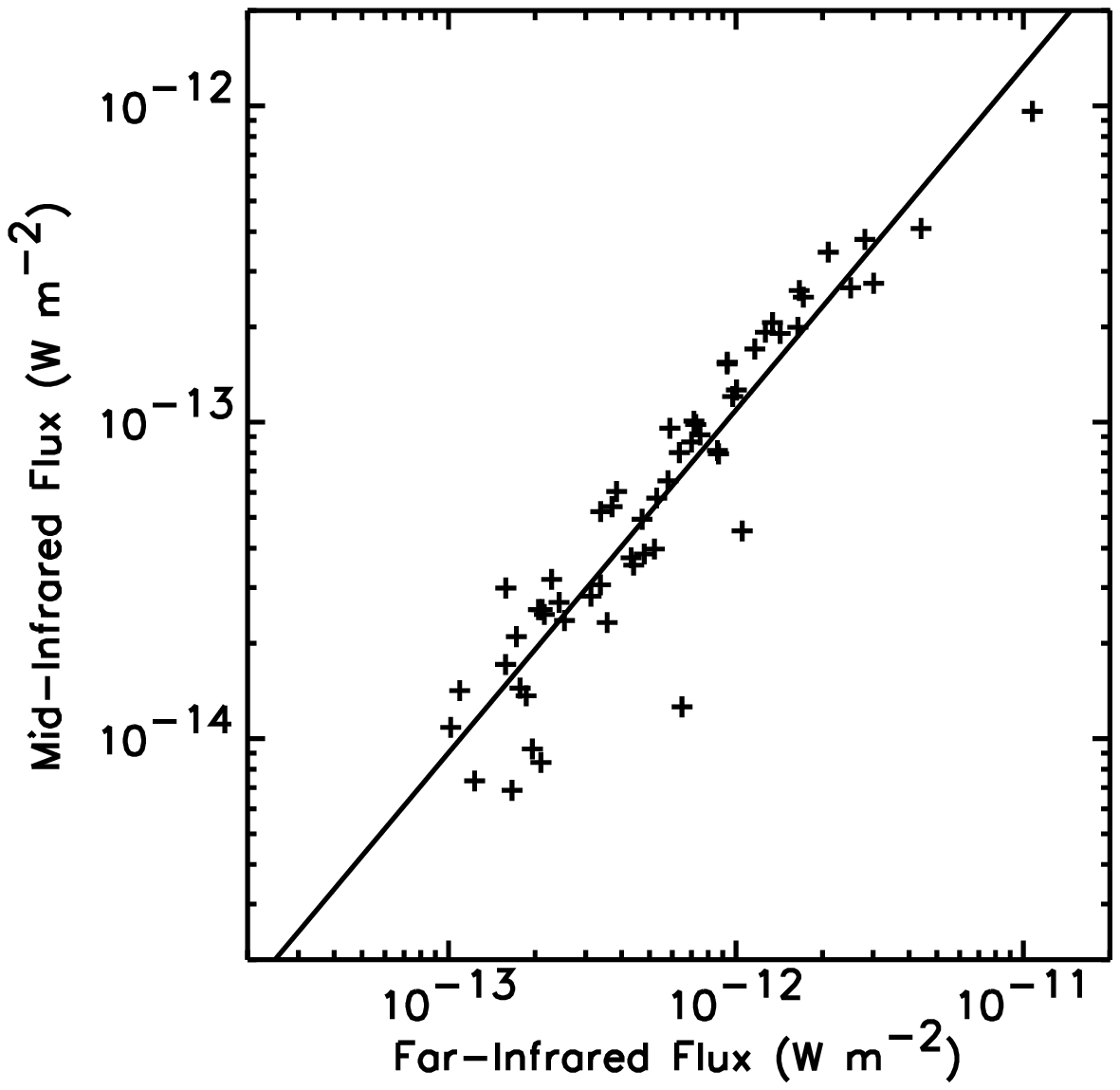,width=3.5in}
\caption{ Plot of the mid-infrared fluxes within $135\arcsec$ versus the
far-infrared fluxes within $135\arcsec$ for all galaxies detected in both
wavebands.  The line shows the best fitting power law.}
\label{f_mirfir}
\end{figure}

We seek a diagnostic of recent star formation activity. Both the mid- and
far-infrared fluxes represent radiation from dust grains heated by
the total stellar radiation field comprised of the
diffuse interstellar radiation field (ISRF) which is dominated by the old
stellar population, and radiation from recent star formation activity.  The
12~$\mu$m data includes both PAH specral line emission and very small grain
continuum emission, both heated by the radiation field.  While the PAH
features will remain relatively unchanging in intensity (or decrease
slightly) as the radiation field increases, the continuum does increase, as
demonstrated by \citet{detal01}.  The 12~$\mu$m data thus includes
contributions from both
the old stellar population, through the diffuse ISRF, and dust heated by
recent star formation.  Furthermore, with the exception of
some Sd and Sm galaxies, the mid-infrared and far-infrared fluxes are
tightly correlated among the galaxies in the sample, as shown in 
Figure~\ref{f_mirfir}.  In particular, the mean
ratios of the mid-infrared flux to the far-infrared flux in Sa, Sb, and Sc
galaxies are statistically identical, as shown in Table~\ref{t_mirfirstat}.  
This correlation demonstrates that mid-infrared fluxes are as effective as the
far-infrared fluxes in tracing the total stellar radiation field.

On the other hand, the K-band flux largely traces the old stellar population,
whose thermal emission peaks at $\sim 1.6 ~\mu$m. Normalizing the 12~$\mu$m
flux densities by the flux densities at K therefore largely cancels out the
contribution from the old stellar population to the radiation field, so that
the 12~$\mu$m/K-band flux density ratio is a good diagnostic of recent star
formation activity.
Similarly, the far-infrared flux densities contain contributions
from dust heated by both the diffuse ISRF and from recent star formation
activity.  Dividing the far-infrared flux densities by the K-band
flux densities tends to cancel the contribution from the diffuse ISRF, thus
making far-infrared/K flux density ratios a second diagnostic for recent star
formation activity.  Therefore, we use the mid- and far-infrared fluxes
normalized by the K-band fluxes as star formation indicators.

To verify that these indicators are useful, we compare these flux ratios
generated from the Starburst99 model \citep{letal99}.  We ran two test cases
with the Starburst99 model: an instantaneous burst of star formation that
forms a fixed mass of stars, and a continuous burst of star formation that
forms stars at a fixed rate.  For each case, we used a Salpeter initial mass
function with a slope of -2.35, an upper mass cutoff of  100~M$_\odot$, and
solar metallicities.  The other inputs that we used for the model match the
defaults for Starburst99.  We examine the diagnostic M$_{Bol}$~-~M$_K$,
effectively the total bolometric flux divided by the K-band flux, which
behaves like our normalized mid- and far-infrared fluxes.  For comparison,
we also examine M$_{Bol}$~-~M$_B$, or bolometric fluxes divided by B-band
fluxes. This is akin to the far-infrared fluxes normalized with B-band
fluxes that are often used in studies that examine trends in star formation
along the Hubble sequence.

\begin{figure}
\psfig{file=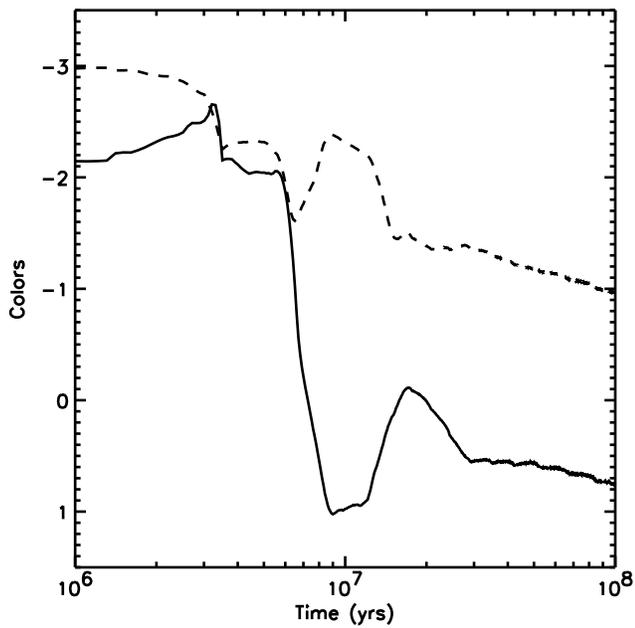,width=3.5in}
\caption{Plots of two star formation
indicators following an instantaneous burst of star formation, as modeled
by Starburst99.  The solid line represents M$_{Bol}$~- M$_K$ and the
dashed line represents M$_{Bol}$~- M$_B$.}
\label{f_modelinst}
\end{figure}

\begin{figure}
\psfig{file=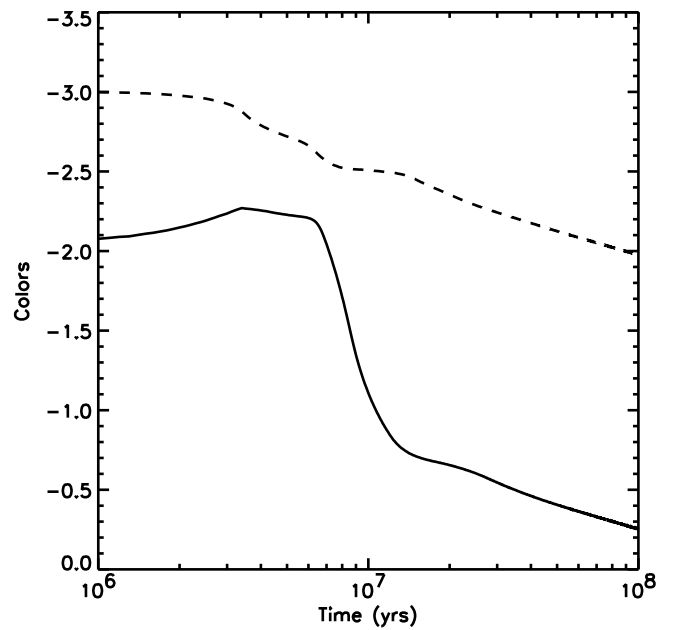,width=3.5in}
\caption{Plots of two star formation indicators following the onset of
continuous star formation, as modeled by Starburst99.  The solid line
represents M$_{Bol}$~- M$_K$ and the dashed line represents M$_{Bol}$~- M$_B$.}
\label{f_modelcont}
\end{figure}

In Figure~\ref{f_modelinst} we plot M$_{Bol}$~-~M$_K$ and
M$_{Bol}$~-~M$_B$ for the time range between 10$^6$ and 10$^8$~yr for an
instantaneous burst of star formation, and in Figure~\ref{f_modelcont} the
same data are plotted for the continuous star formation model.  The
colors on the left side of the graphs represent younger stellar populations,
while the colors on the right side of the graphs represent older stellar
populations.  In both plots, the most notable feature is the dip in the
M$_{Bol}$~-~M$_K$ indicator at 10$^7$~yr due to the evolution of the most
massive stars into the red supergiants that dominate the total
K-band emission at that time.  It is evident that, for both scenarios, the
difference between recent star formation and an evolved
stellar population for M$_{Bol}$~-~M$_K$ are 1.5 - 2.5 magnitudes, and the
K-band normalization provides one magnitude increased discrimination
compared to normalization by B-band fluxes.  Since the mid- and far-infrared
fluxes are proportional to the bolometric luminosity, diagnostics using
these fluxes divided by K-band fluxes are useful for tracing recent star
formation, and are certainly preferable to diagnostics using  B-band fluxes.
Real galaxies will of course contain star formation regions lying
within populations of older stars, and star formation activity may occur in a
series of instantaneous bursts, or be continuous but time-varying.  What
this exercise does show is that the diagnostics we have chosen are effective
for discriminating recent star formation activity.

\section{Star Formation Along the Hubble Sequence}

\subsection{Star Formation Activity}

We will examine trends in star formation along the Hubble sequence in three
regions of these galaxies.  The global regions were defined as the regions
within the inner 135~arcsec because it matched the aperture of ISOPHOT
observations at 60 and 100~$\mu$m.  The nuclear regions were defined as the
inner 15~arcsec, which corresponds to a diameter in an integer number of
pixels that is wider than three times the full width
half-maximum of the point spread function in the ISOCAM images.  The disk
regions were defined as outside the nuclear regions but within the global
regions.  To study the global
star formation, we use the  far-infrared fluxes (the 40~- 220~$\mu$m
fluxes from Paper 1) normalized by the K-band fluxes for 63 galaxies where
we measured both fluxes within  135~arcsec.  To study the nuclear and disk
star formation, we use the mid-infrared (8-15~$\mu$m) fluxes
normalized by the K-band fluxes.  Nuclear star formation was investigated in
70 galaxies in which fluxes were detected within 15~arcsec at $\ge$ 3~$\sigma$
levels, and disk star formation was explored in 55 galaxies in which fluxes
within 135~arcsec were detected at $\ge$ 3~$\sigma$.  The use of
fixed angular apertures means that the physical areas sampled in each
galaxy will vary with the distance to the galaxy.  However, the flux ratios
used here are uncorrelated with distance, so differences in the
physical area sampled should not signifantly affect the analysis.

\begin{figure}
\psfig{file=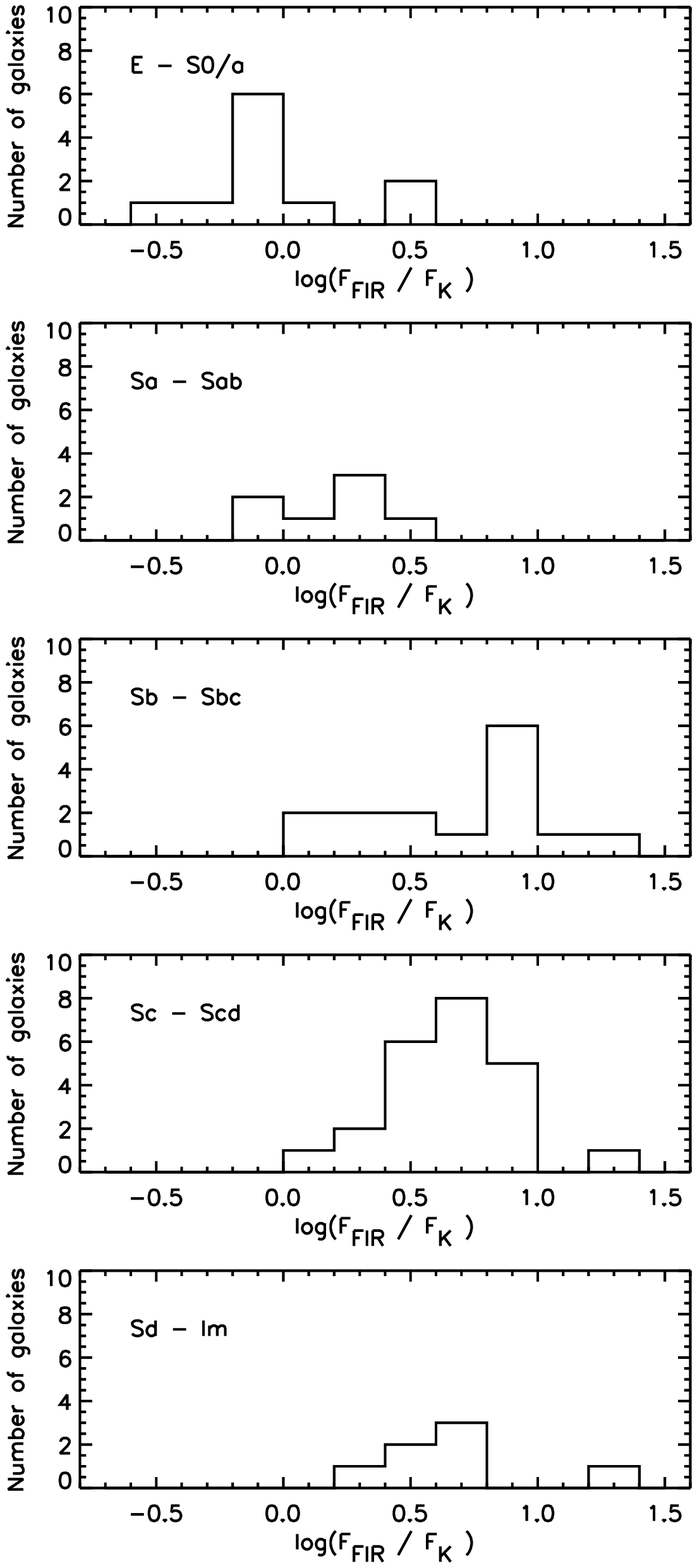,width=3.5in}
\caption{Histograms of the global star formation diagnostic
log($\frac{F_{FIR}}{F_K}$) for subsets of the galaxies in the sample binned
according to Hubble type.}
\label{f_firglobsfr}
\end{figure}

\begin{figure}
\psfig{file=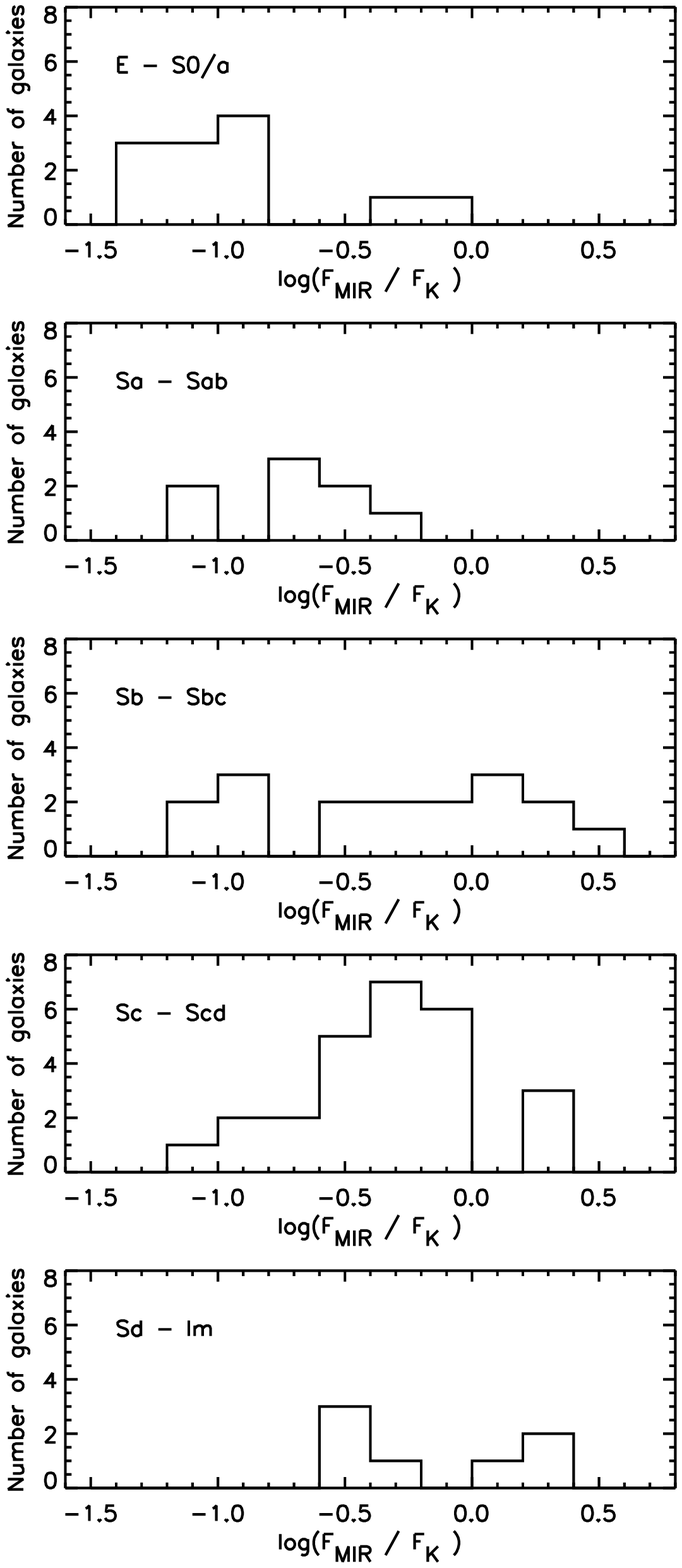,width=3.5in}
\caption{Histograms of the nuclear star formation diagnostic
log($\frac{F_{MIR}}{F_K}$) for subsets of the galaxies in the sample binned
according to Hubble type.}
\label{f_mirnucsfr}
\end{figure}

\begin{figure}
\psfig{file=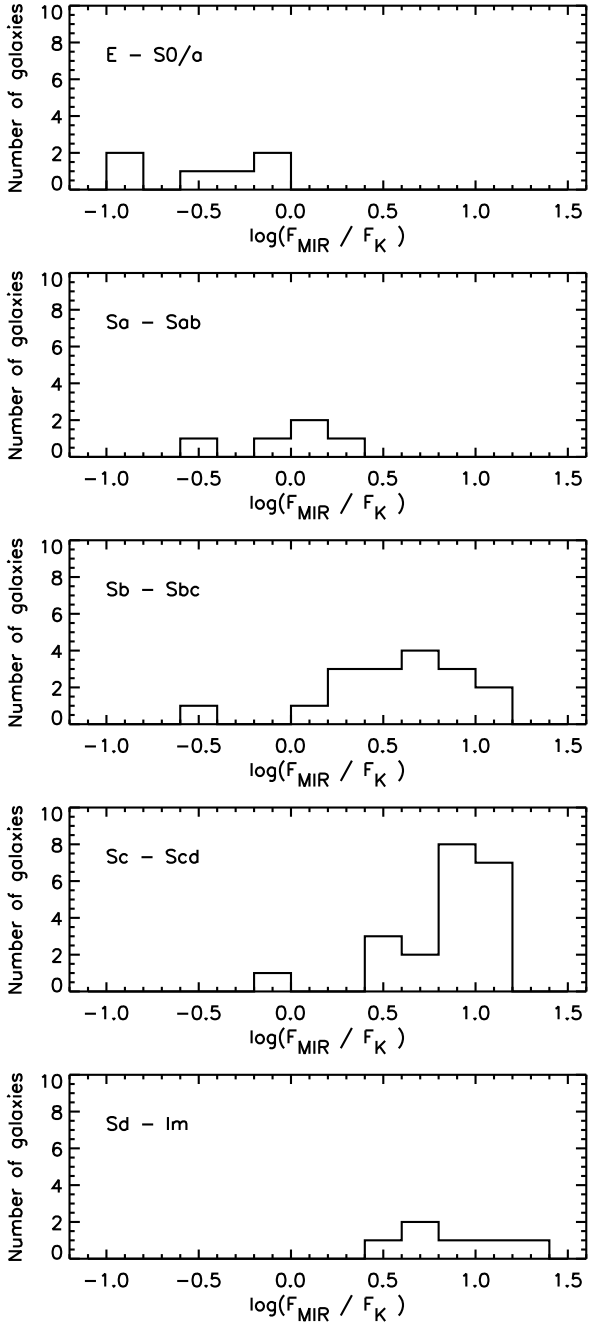,width=3.5in}
\caption{Histograms of the disk star formation diagnostic
log($\frac{F_{MIR}}{F_K}$) for subsets of the galaxies in the sample binned
according to Hubble type.}
\label{f_mirdisksfr}
\end{figure}

Figures~\ref{f_firglobsfr}~- \ref{f_mirdisksfr} show histograms of the star
formation indicators for each Hubble type.  A clear trend can
be seen in these histograms, with star formation activity increasing from
early-type to late-type galaxies.  This trend is seen separately in global,
nuclear, and disk star formation, although it is not as well defined in the
nuclear regions as it is in the disk regions.

Table~\ref{t_sfrstat} presents Gaussian statistics that compare the star
formation indicators of each Hubble type.  Clearly these statistics
   show quantitatively a large difference in the star formation activity
between Sa~- Sab and Sc~- Scd galaxies.  The difference between mean
   values of the star formation diagnostics for Sa~- Sab and Sc~- Scd galaxies
is greater than 3~$\sigma$ for all regions, and the difference is nearly
6~$\sigma$ for the disk regions.  Because the data may not necessarily obey
Gaussian statistics, we also used non-parametric statistical methods to
compare Sa~- Sab, Sb~- Sbc, and Sc~- Scd galaxies.  Table~\ref{t_sfrnonpar}
contains results from applying the Kolmogorov-Smirnov (K-S) test to the
diagnostics for two different morphological subsets.  The probabilities
reported indicate the likelihood that the two populations come from the same
distribution (\textit{i.e.} they have similar values).    The
K-S tests show at the 98~\% confidence level that the Sa~- Sab and Sc~- Scd
nuclear star formation diagnostics have distinct values and at better than
the 99~\% confidence level that the Sa~- Sab and Sc~- Scd global and disk
star formation diagnostics have distinct values.

It has been suggested that use of RSA Catalog classifications for studies such
as these can produce spurious correlations, since star formation is included in
the classification scheme.  The analysis above has used RC3 classifications,
but we have repeated the analysis using RSA classifications.  We found no
changes in the trends or their statistical significance for nuclear or global
star formation, but the correlation of disk activity with morphological
type went from $\sim$~6~$\sigma$ to $\sim$~8~$\sigma$.  Evidently the star
formation features do make a measurable difference in the RSA
classifications.

These data all clearly show that star formation activity increases
along the Hubble sequence from early- to late-type galaxies, with a  strong
distinction between the star formation activity of Sa~- Sab and Sc~- Scd
galaxies.  The trend is clear for the global, nuclear, and disk regions,
although it is not as well defined in the nuclear regions.

\subsection{Spatial Extent of Star Formation}

In Paper 1 we noted that late-type galaxies usually contained well
defined star formation structures in their disks while early-type galaxies
often did not.  Furthermore, some late-type galaxies hosted their strongest
sites of star formation in their disks, whereas all early-type galaxies
produced their strongest mid-infrared emission from within or near their
nuclei.  Our qualitative conclusion was that star formation is more
centralized in early-type galaxies and more extended in late-type galaxies.

We now attempt to show this trend quantitatively using the 12~$\mu$m
mid-infrared fluxes.  To indicate the central concentration of the star
formation activity,  we calculate the logarithm of the flux within
15~arcsec divided by the  flux within 135~arcsec, or log
($\frac{F_{MIR}(15\arcsec)}{F_{MIR}(135\arcsec)}$). This quantity will be
larger for a centrally concentrated galaxy and smaller for a more extended
galaxy. (Taking the logarithm of this quantity is appropriate because we are
looking for multiplicative differences in the ratio, which will  appear
linear in logarithm space.)  We perform this analysis for only for those
galaxies where the 12~$\mu$m flux within the  135~arcsec aperture was
detected at the 3~$\sigma$ level. Because this quantity  may also be related
to the angular width of the galaxy examined, we  only use galaxies with
similar angular sizes, those with log (D$_{25}$) values (from RC3) between
1.5 and 1.9.  These constraints pared the sample down to 44 galaxies.

\begin{figure}
\psfig{file=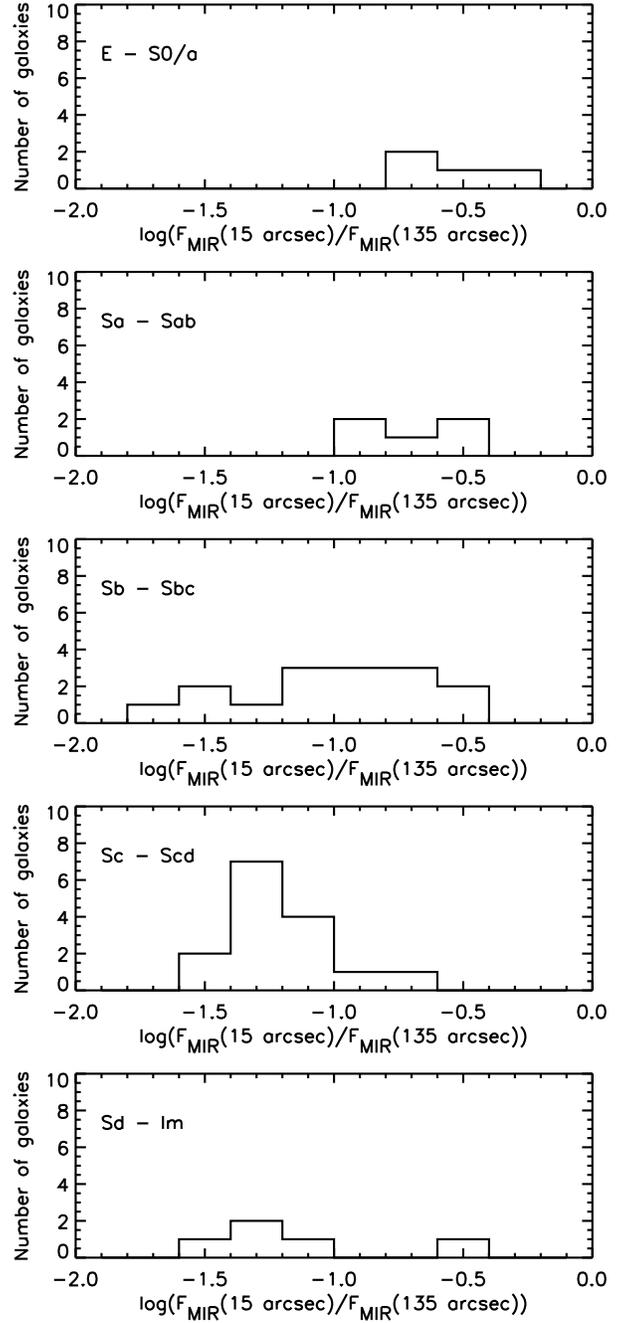,width=3.5in}
\caption{Histograms of the diagnostic of the central concentration of star
formation, log ($\frac{F_{MIR}(15\arcsec)}{F_{MIR}(135\arcsec)}$),
for subsets of the galaxies in the sample binned
according to Hubble type.  Only galaxies where
1.5~$<$ log(D$_{25}$)~$<$ 1.9 were used in these histograms.
The diagnostic is high for galaxies that are centrally concentrated and low
for galaxies that are extended.}
\label{f_mirrat}
\end{figure}

Table~\ref{t_mirrat} presents Gaussian statistics on the
log ($\frac{F_{MIR}(15\arcsec)}{F_{MIR}(135\arcsec)}$) values, and
Figure~\ref{f_mirrat} presents histograms showing the distribution of values
for each Hubble type.  Since this data is surely non-Gaussian,
we also present probabilities resulting from the K-S test in
Table~\ref{t_mirratnonpar}.  These data show a continuous trend from Sa to
Scd with a 5~$\sigma$ significance.  Alternatively, the K-S test shows at the
99\% level that the Sa-Sab and the Sc-Scd galaxies differ in the spatial
extent of their mid-infrared fluxes.

\subsection{Discussion}

Perhaps ironically, the ISO Atlas results are consistent with nearly all
previous optical and ultraviolet studies - including \citet{s82}, \citet{kk83},
\citet{k83}, \citet{detal87}, \citet{ktc94}, and \citet{detal94} - but
inconsistent with nearly
all the earlier infrared photometric studies. It is intriguing that our
infrared approach has
confirmed the trends with Hubble type found in optical studies.  Does this mean
that the issues of extinction mentioned in the Introduction are actually
irrelevant?  We suggest that extinction is an issue, and the infrared approach
provides a statistically stronger conclusion.  Unfortunately, it is not
possible to make quantitative comparisons because none of the
optical studies that reported a trend with Hubble type offered any
quantitative assessment of confidence.  Most simply showed the global trend
graphically as we have done in Figures~\ref{f_firglobsfr}~- \ref{f_mirdisksfr}.

The far-infrared spectroscopic results from \citet{lvhetal99} are consistent 
with the ISO Atlas results, but the only two far-infrared photometric studies 
with results consistent with these are
\citet{dbs87} and \citet{pr89}.  Both these studies used RSA morphological
classifications, and as we have argued above, trends found in samples
classified with the RSA Catalog will at least partly reflect the
classification criteria rather than true physical trends.   The remaining
photometric infrared studies, including \citet{setal83},
\citet{detal84}, \citet{dy91}, \citet{if92}, \citet{tts96}, and \citet{dh97}
found no significant correlation between star formation activity and Hubble
type.  We suggest that the advantages the ISO measurements provide, both the
angular resolution and the point-and-integrate observing mode, compared to
IRAS, and the normalization by K-band fluxes rather than by B-band fluxes,
are responsible for the different conclusions reached in this study compared
to most of the previous infrared studies.

These trends in star formation must reflect physical differences between early-
and late-type spiral galaxies.  These physical differences can be
interpreted in
the context of the results from \citet{k89}, which states that star formation
is related to gas density by a simple power law and that the gas density
must exceed a threshold density before star formation is triggered.  This must
mean that the gas density is more likely to exceed the critical threshold in
late-type spiral galaxies than in early-type spirals.  In fact, the molecular
and total gas surface  density does increase from early- to late-type galaxies
\citep{yk89}, \citep{rh94}, and the ratio of total gas mass to dynamical mass
also increases from early- to late-type galaxies \citep{ys91}.  This
demonstrates that a  typical Sc galaxy is more likely than an Sa galaxy to have
gas densities  high enough to trigger gravitational collapse and star
formation.   Interestingly, \citet{yetal95} also demonstrate that molecular gas
is  distributed over a larger area in late-type spiral galaxies than in
early-type spiral galaxies.  This suggests that our observations of more
spatially extended star formation in late-type galaxies compared to
early-type galaxies may be the result of the
spatial distribution of molecular gas in spiral galaxies.

\section{Dependence of Star Formation on Other Factors}

\subsection{Relation of Nuclear Star Formation to Bars}

\subsubsection{Background}

Unlike research on variation in star formation along the Hubble sequence,
research on how bars may enhance star formation is reaching a consensus, but
the paradigm is more complicated than for star formation along the Hubble
sequence.

Very early studies on barred galaxies, such as \citet{sp65} and
\citet{sp67}, demonstrated that barred galaxies are more likely to
host enhanced
nuclear star formation. Studies using IRAS data, including
\citet{detal84} and \citet{hetal86}, demonstrated that barred galaxies have
enhanced star formation compared to unbarred galaxies.  These results have
been confirmed  by \citet{d87}, \citet{hetal96}, \citet{hfs97a}, and
\citet{retal01}, who
find the enhancement only in early-type spiral galaxies.  These results are
not confirmed
by \citet{if92}, but their study did not subdivide the data into early- and
late-type spiral galaxies, so they may not have been able to find the effect
seen by others.

Although these studies tend to agree, most of them rely on IRAS data to detect
nuclear enhancements in star formation, even though the IRAS detectors
may not adequately isolate nuclear infrared emission from disk
emission.  Because the data in the ISO Atlas has several advantages in
studying star formation, described earlier, we attempt to examine the issue
of whether bars enhance star formation.  Unfortunately, this sample contains
very few early-type galaxies, where the effect is expected to be seen, so we
cannot expect to produce a definitive conclusion on this subject.

\subsubsection{Analysis}

In this analysis, we compare SA, SAB, and SB galaxies for three subsets
of galaxies in the sample: Sa~- Sab galaxies, Sb~- Sbc galaxies, and Sc~- Scd
galaxies.  Since bars specifically affect nuclear star formation,
we use the nuclear star formation indicator log($\frac{F_{MIR}}{F_K}$).
Table~\ref{t_mirnucsfr_bar} presents Gaussian statistics on the
log($\frac{F_{MIR}}{F_K}$) values.  We also present probabilities from
applying  the K-S test in Table~\ref{t_mirnucnonpar_bar}.

The Gaussian statistics show very little statistical difference between the
SA, SAB, and SB galaxies for any subset.  For Sa~- Sab galaxies,
we have very few galaxies for making a comparison, so more data would be
needed to properly determine the presence or absence of a trend.  For Sb~- Sbc
and  Sc~- Scd galaxies, the number of galaxies in the comparison is
large enough that the statistics can be treated as meaningful.  No difference
is found between the SA and SB types for Sb~- Sbc galaxies.  The data do
show an increase in star formation from SA to SB galaxies for Sc~- Scd
galaxies, but the trend is a statistically weak 2~$\sigma$ trend.  The
application of the K-S test to  the data for all galaxy subsets confirms that
there is no significant statistical difference between SA and SB galaxies.

\subsubsection{Discussion}

The current paradigm suggests that only in some barred galaxies will enhanced
star formation be found, and then only in early-type spiral galaxies.  The
ISO Atlas sample contains too few SA and SAB early-type spiral galaxies to
make any definitive  statements on how bars may affect nuclear star
formation.  Some late-type  spiral galaxies apparently host enhanced star
formation, but the trend is  statistically weak, so we cannot make any
definitive statements.  Clearly,  a larger sample of galaxies is required to
make stronger statements on the  effects of bars.  We can however
express agreement with most other  recent results that demonstrate little
enhancement of nuclear star formation  in barred late-type galaxies.

\subsection{Relation of Nuclear Star Formation Activity to AGN Activity}

\subsubsection{Background}

The association of starbursts with active galactic nuclei (AGNs)
has been postulated for both Seyferts and low ionization nuclear emission
regions (LINERs).  The issues for these two classes of galaxies are different,
however.

For Seyferts, the issue is how enhanced star formation activity is connected
to the AGN activity.  A few studies, such as \citet{rrj87}, \citet{dmm88},
\citet{hetal89}, \citet{gp93}, \citet{cetal97}, \citet{getal98}, and
\citet{oetal99} have found links between the Seyfert activity and star
formation activity, although most suggest that the enhancement in star
formation is seen only in Seyfert 2 galaxies.  Since the ISO Atlas has
superior spatial resolution at mid-infrared wavelengths compared to most other
previous infrared studies of the Seyfert/starburst connection, we should be
able to examine whether enhanced star formation activity is found in Seyferts.
Unfortunately, using the  \citet{hfs97b} classifications for AGN, the sample
contains only three Seyfert galaxies.

For LINERs, the issue is whether or not LINER optical line emission is
correlated with starburst activity.  Three types of excitation mechanisms have
been proposed for LINER emission: photoionization from  a cluster of hot young
stars \citep{tm85, s92}, photoionization from a  power-law source such as an
AGN \citep{hfs93}, and shock excitation \citep{h80}.  It is not clear that
only one excitation mechanism is responsible for producing all observed LINER
activity, as recent studies have  suggested that the LINER class of galaxies
may consist of galaxies with  different central engines \citep{letal98}
\citep{aetal00}. If LINER emission is powered by photoionization by hot young
stars from a  starburst, the nuclei should produce relatively strong
mid-infrared emission. If LINER emission is powered by other mechanisms,
however, the LINER nuclei may not necessarily be sources of strong mid-infrared
emission.

To address these issues, we compare the nuclear star formation activity
in Seyferts, LINERs, H~{\small II} galaxies, and transition objects in our
sample that have also been surveyed by \citet{hfs97b}, who classified the
galaxies' nuclear activity based strictly on optical spectral line ratios.
(NGC~6503 is classified as both a transition object and a Seyfert,
but we treat it as a transition object, which is, according to \citet{hfs97b},
the more probable designation for the nuclear activity.)  By using this one
source of AGN classification, we ensure that the classifications are consistent
among the galaxies in the comparison.

\subsubsection{Analysis}

\begin{figure}
\psfig{file=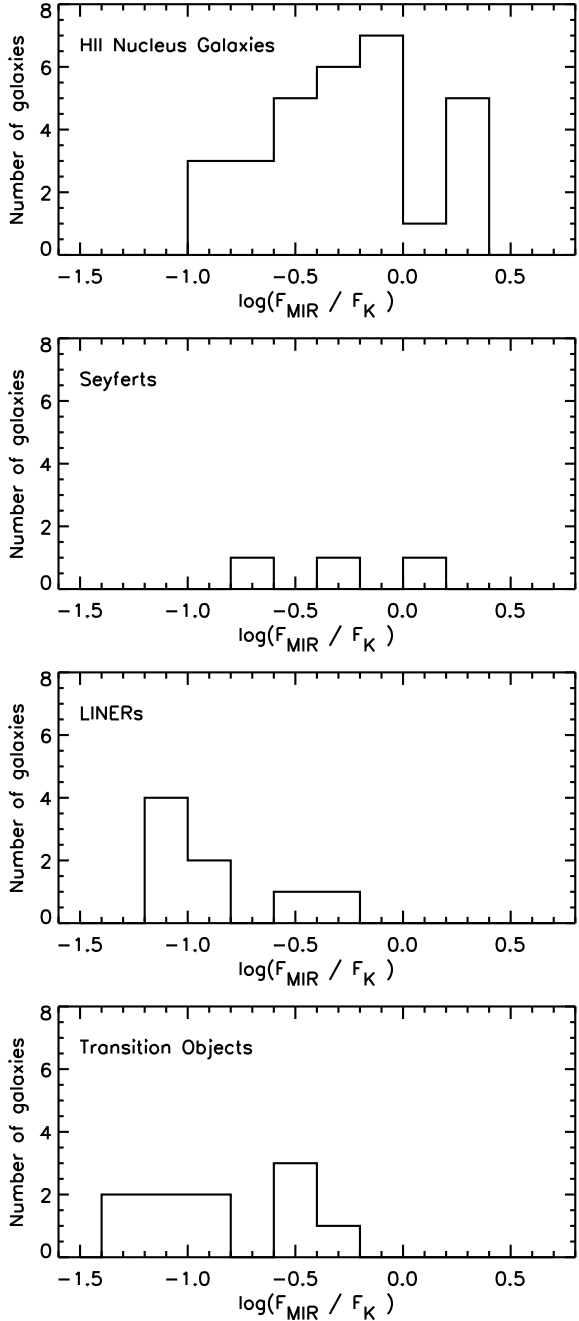,width=3.5in}
\caption{Histograms of the nuclear star formation diagnostic
log($\frac{F_{MIR}}{F_K}$) for subsets of the galaxies in the sample based
on the nuclear activity classifications in \citet{hfs97b}.  Note the
relatively lower star formation activity in LINERs and transition objects
compared to H~{\small II} galaxies and Seyferts.}
\label{f_mirnucsfr_agn}
\end{figure}

We compare the various classes of galaxies using the nuclear star formation
indicator log($\frac{F_{MIR}}{F_K}$), where the fluxes are measured for the
inner 15~arcsec. Table~\ref{t_mirnucstat_agn}  presents Gaussian statistics
on the log($\frac{F_{MIR}}{F_K}$) values, and Figure~\ref{f_mirnucsfr_agn}
presents histograms showing the distribution of  values for each Hubble type.
We also present probabilities from applying the  K-S test in
Table~\ref{t_mirnucnonpar_agn}.

It is very apparent from the data that the Seyferts and H~{\small II} galaxies
are sites of stronger mid-infrared activity than the LINERs and
transition objects.  The LINERs and transition objects also seem to be very
similar to each other in terms of nuclear mid-infrared activity, which may
indicate that similar physical processes are responsible for the generation
of the spectral line emission in both these classes of galaxies.
The Gaussian statistics show that the H~{\small II} galaxies and Seyferts
have higher log($\frac{F_{MIR}}{F_K}$) values than the LINERs and transition
objects.  The statistical difference between the H~{\small II} galaxies and
the LINERS and  transition objects is $\sim$~6~$\sigma$. The results from the
K-S test show that the  H~{\small II} galaxies have a distinctly different
distribution of values than the LINERs or transition objects at the 99~\%
confidence level.

\subsubsection{Discussion}

These results demonstrate very conclusively that the LINERs and transition
objects in this sample are, at the very least, not classic starburst
galaxies.  They could be either quiescent AGN, or galaxies with
shock-dominated nuclear regions.  The contrast in mid-infrared emission
between the more active Seyferts and  the quiescent LINERs is also very
striking, although the number of Seyferts in this sample is small.  The data
may imply that the presence of either circumnuclear material or
circumnuclear star formation will boost the AGN activity from the more
quiescent LINER level to the more active Seyfert level. However, more work
is needed to investigate the connection, specifically a more careful study
of the circumnuclear regions of Seyferts and LINERs as well as a
mid-infrared survey of all Seyferts and LINERs from the sample in
\citet{hfs97b}.

It is noteworthy, but perhaps not surprising, that the H~{\small II} galaxies
and Seyferts are indistinguishable using these photometric diagnostics.  This
is, after all, the underlying cause of the "starbursts-monsters" controversy
\citep{hetal83}.

\section{Star Formation Rates and Molecular Gas Content}

\subsection{Global Star Formation Rates}

To measure the range of star formation activity in the ISO Atlas sample, we
have converted the far-infrared fluxes into quantitative star
formation rates by
assuming that all the far-infrared flux comes from star formation regions.
We use the formula
\begin{equation}
SFR(M_\sun yr^{-1}) = 1.76 \times 10^{-10} L_{FIR} (L_\sun)
\end{equation}
where $L_{FIR}$ is the 8~- 1000~$\mu$m flux \citep{k98b}.  Both
\citet{st92} and
\citet{k98a} have emphasized that direct conversion of far-infrared fluxes to
star formation rates relies on the assumptions that far-infrared fluxes are
emitted from dust heated by star formation regions and that the dust is
optically thick.  Dust heated by  an interstellar radiation field generated by
old stars will also produce far-infrared emission, leading to star formation
rates that are calculated to be higher than they really are.  Nonetheless, we
will use this conversion factor as an estimate of the global star formation
rates in these galaxies.

\begin{figure}
\psfig{file=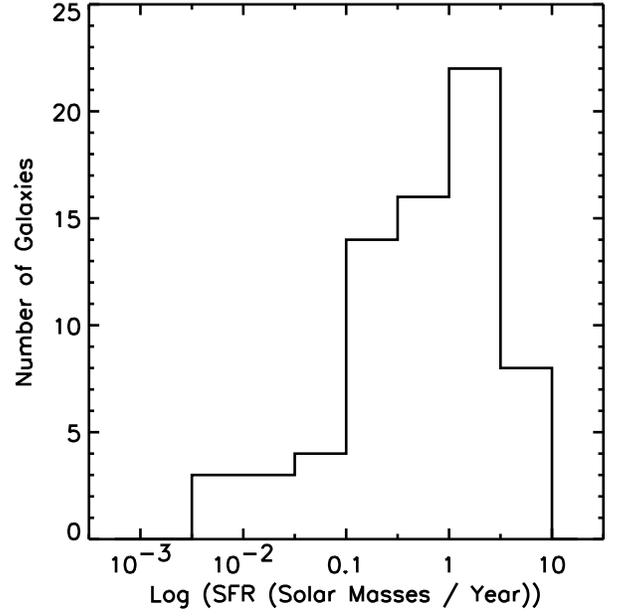,width=3.5in}
\caption{Histogram of the distribution of star formation rates for all
galaxies in the sample where far-infrared luminosities were calculated.}
\label{f_sfrhist}
\end{figure}

First we converted the global 40~- 220~$\mu$m fluxes (\textit{i.e.} the fluxes
from within the inner 135~arcsec) to 8~- 1000~$\mu$m fluxes using the
conversion factor 1.192 $\pm$ 0.235.  Following a method similar to the
one described in \citet{setal00}, this conversion factor and its
uncertanties were calculated  by assuming that the emission in the 8~-
1000~$\mu$m range is dominated by a  single blackbody component with
temperatures ranging between 20~- 40~K and emissivities with wavelength
dependence ranging from $\lambda^0$ to
$\lambda^{-2}$. Then luminosities were calculated, and the above formula was
applied to convert the luminosities into star formation rates.
Figure~\ref{f_sfrhist} shows the range of star formation rates.  The average
star formation rate is 1.4~M$_\sun$~yr$^{-1}$, with a maximum of
8.5~M$_\sun$~yr$^{-1}$ and a minimum of 0.0035~M$_\sun$~yr$^{-1}$.  These
results are affected by Malmquist bias because the sample was
magnitude-limited. Also, the validity of the conversion factor has not been
tested through observations. Nonetheless, these star formation rates are
consistent with the spiral disk star formation rates listed in \citet{k98a}.

\subsection{Gas Consumption Times}

Since we have estimated global star formation rates for these galaxies we
should also examine whether the galaxies contain sufficient gas to maintain
this level of star formation activity over a Hubble time. This issue has
been investigated previously by  \citet{ltc80}, \citet{k83},
\citet{s86}, \citet{detal87}, \citet{ktc94}, and \citet{dh97}.  All of these
studies have found gas consumption times of a few times 10$^9$~yr, which is
shorter than a Hubble time.  They usually suggest that gas recycling
is responsible for increasing the gas consumption time, although only
\citet{ktc94} have modeled this effect.  However, most of the above studies
have relied on optical or ultraviolet data for determining star formation
rates.  Dust extinction in these wavebands may lead to calculated star
formation rates that are artificially low, which then may lead to calculated
gas consumption times that are artificially high.  We attempt to address this
issue with our far-infrared data, despite the concerns about using the data
to calculate star formation rates mentioned above.

To calculate gas consumption times, we will use the ``Roberts time''
\begin{equation}
\tau_R = \frac{M_{gas}}{SFR}
\end{equation}
suggested by \citet{ktc94}.  For gas masses, we used the CO data from
the FCRAO Extragalactic CO Survey \citep{yetal95} to estimate the gas masses
within 135~arcsec apertures.
We calculate gas consumption times for the 20 galaxies in the ISO sample
for which the CO intensities were detected at the 3~$\sigma$
level in the central beam.
The average gas consumption time we  calculate is 6~$\times$~10$^{8}$~yr,
with a maximum of
1.3~$\times$~10$^{9}$~yr and a minimum of 3~$\times$~10$^{8}$~yr.
These gas consumption times seem to be independent of Hubble type
or the star formation metrics we used above.

The gas consumption times we calculate are short compared to the
few times 10$^9$~yr values found in most other studies.   It is possible that
the high star formation rates we calculate are more accurate than those
calculated using data from optical and ultraviolet wavebands because of dust
extinction issues.  Alternatively, our star formation rates could plausibly be
high for several reasons.   There may be contributions to the far-infrared flux
from dust heated by the old stellar population that could increase our
calculated star formation rates.  Also, if we had used a different conversion
factor to convert infrared flux to star formation, for example that given by
\citet{sy83}, we would have found gas consumption times on the order of
10$^9$~yr, which would be more in line with most previous results (most closely
with \citet{dh97}, who also calculated gas consumption times from far-infrared
data).  However, our global star formation rates are comparable to those listed
in Kennicutt's review \citep{k98a}, so our star formation rates may be
correct.  In summary, multiple  factors may be responsible for the slightly
low gas consumption  times we find compared to most previous results,
but because of uncertainties in the assumptions, we cannot argue that our
results are more accurate than most previous calculations.

\subsection{Testing the Schmidt law}

In discussing the trend in star formation with Hubble type, we suggested that
it resulted from the difference in molecular gas content between
early- and late-type galaxies.  If so, for galaxies where we
have both infrared and CO data, we should be able to find the power law
relation between star formation and gas density discussed by \citet{s59} and
\citet{k89}.  We investigate this with the molecular gas mass data from
\citet{yetal95} and do the comparison with star formation activity for
global, nuclear, and disk regions separately.  For the global analysis we
use the star formation rates found above, and for the nuclear and disk
analysis we use the 12 and 60~$\mu$m flux densities, which have spatial
resolutions that are equivalent to or better than the CO data.  Because we were
limited by the resoulution of the CO data, we used 45~arcsec apertures to
represent the nuclear regions (instead
of the 15~arcsec aperutres used in the analysis of star formation trends).

\begin{figure}
\psfig{file=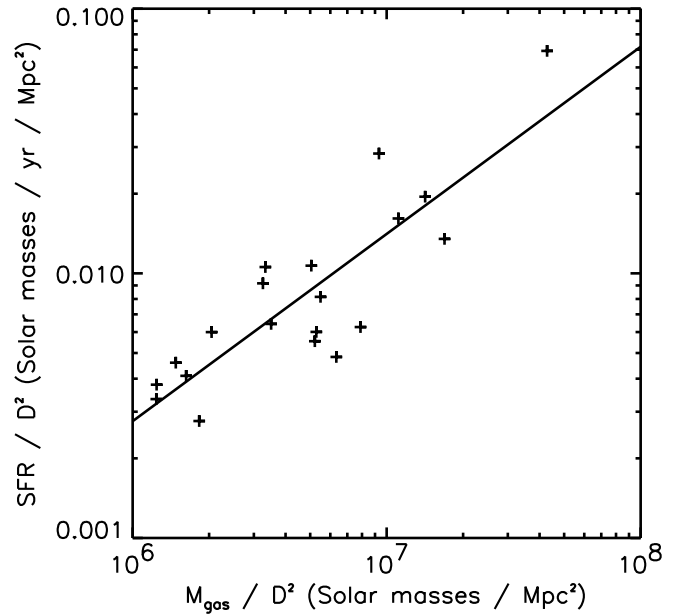,width=3.5in}
\caption{Plot of the global star formation rate versus molecular gas mass
for galaxies in both the ISO and FCRAO samples.  Both quantities have been
normalized by the distance squared for reasons explained in the text.  
The line shows the best fitting power law.}
\label{f_globsch}
\end{figure}

Figure~\ref{f_globsch} shows the gas masses versus star formation rates
normalized by the square of the distance to the galaxies. We use this
normalization because both quantities are effectively derived from
luminosities; otherwise there may be a spurious correlation since one is
effectively plotting D$^2$ vs. D$^2$.  The figure shows that the two
quantities are very well correlated and can be represented by a simple power
law.

\begin{figure}
\psfig{file=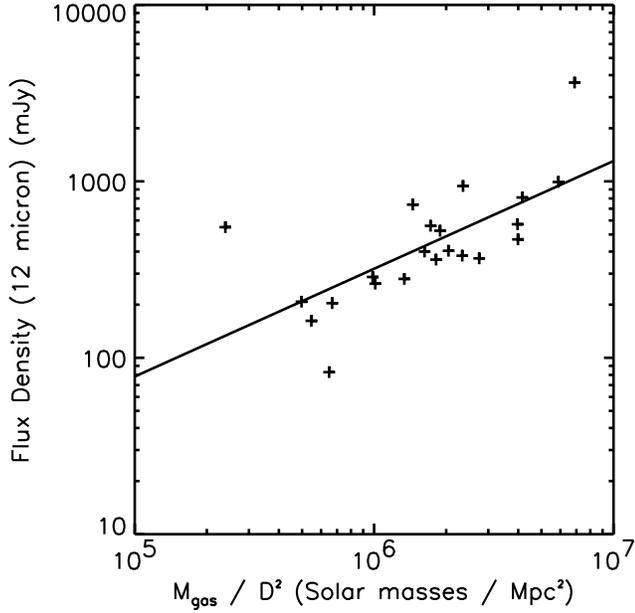,width=3.5in}
\caption{Plot of the 12~$\mu$m flux densities versus molecular gas mass for the
nuclei of galaxies in both the ISO and FCRAO samples.
The gas masses in this figure and Figures~\ref{f_nuc60sch}~-
\ref{f_disk60sch} have been normalized by the distance squared for reasons
explained in the text.  The lines in this figure and Figures~\ref{f_nuc60sch}~-
\ref{f_disk60sch} show the best fitting power laws.}
\label{f_nuc12sch}
\end{figure}

\begin{figure}
\psfig{file=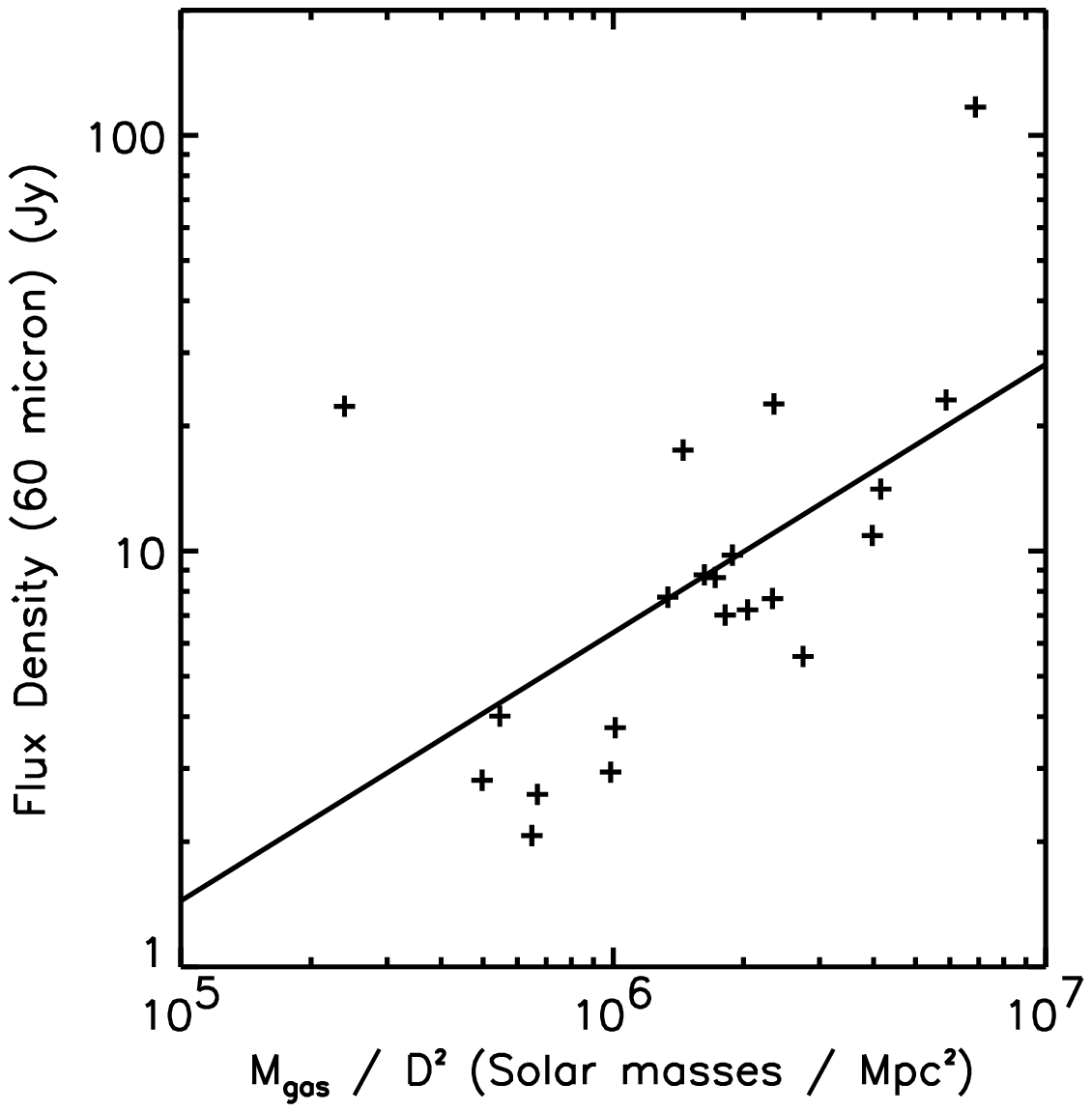,width=3.5in}
\caption{Plot of the 60~$\mu$m flux densities versus molecular gas mass for the
nuclei of galaxies in both the ISO and FCRAO samples.}
\label{f_nuc60sch}
\end{figure}

Figures~\ref{f_nuc12sch} and~\ref{f_nuc60sch} show the normalized gas
masses versus 12 and 60~$\mu$m flux densities
for the nuclear regions of the galaxies.  Most of the galaxies
lie very nicely along this trend.  The exceptions are NGC~1569 (the outlier
near the upper left of the figures) and NGC~5236 (the outlier
near the upper right of the figures).  NGC~1569 is a nearby dwarf galaxy
with strong star formation activity, so it may not necessarily follow the
same trend as most spiral galaxies.

\begin{figure}
\psfig{file=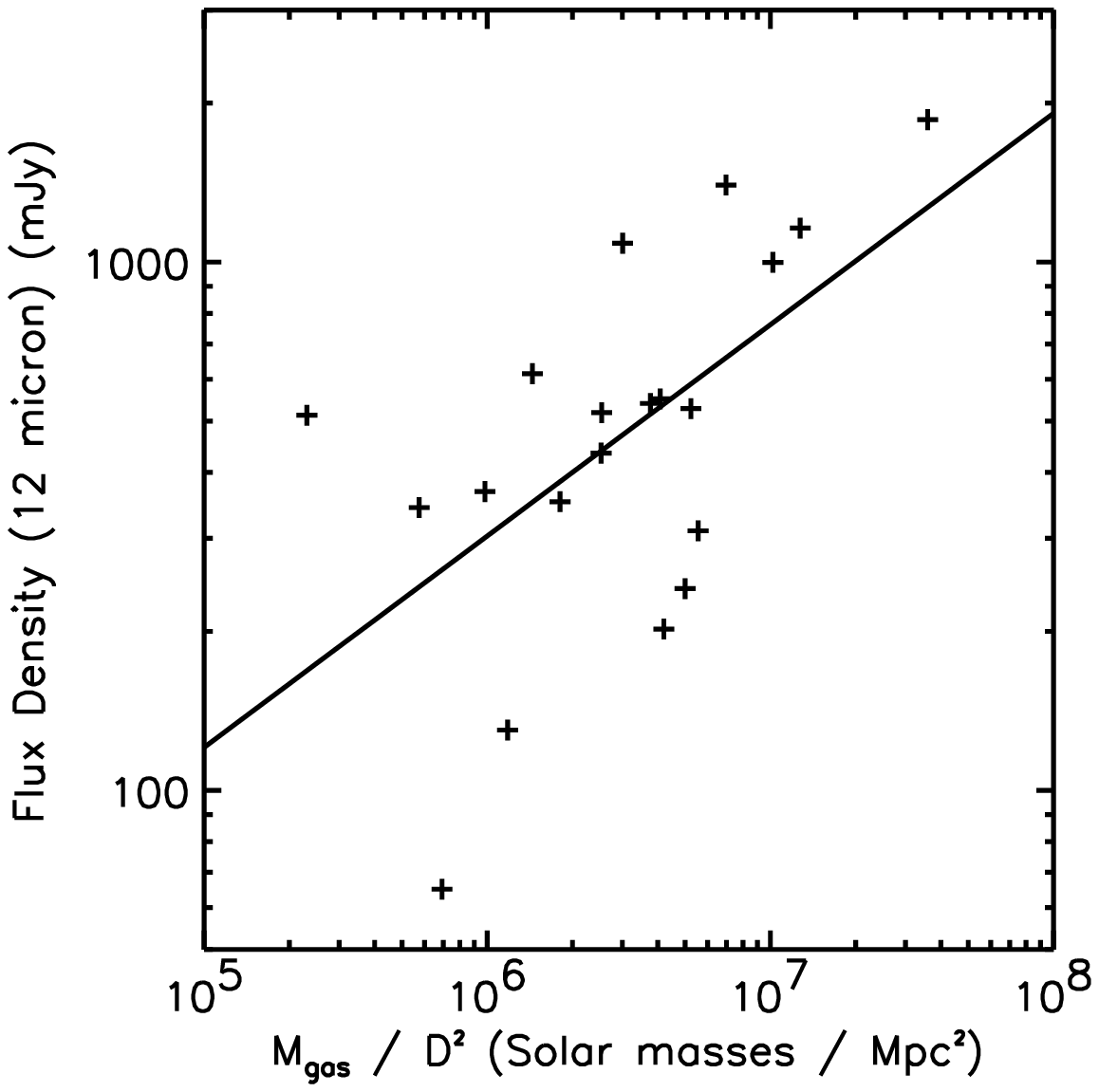,width=3.5in}
\caption{Plot of the 12~$\mu$m flux densities versus molecular gas mass for the
disks of galaxies in both the ISO and FCRAO samples.}
\label{f_disk12sch}
\end{figure}

\begin{figure}
\psfig{file=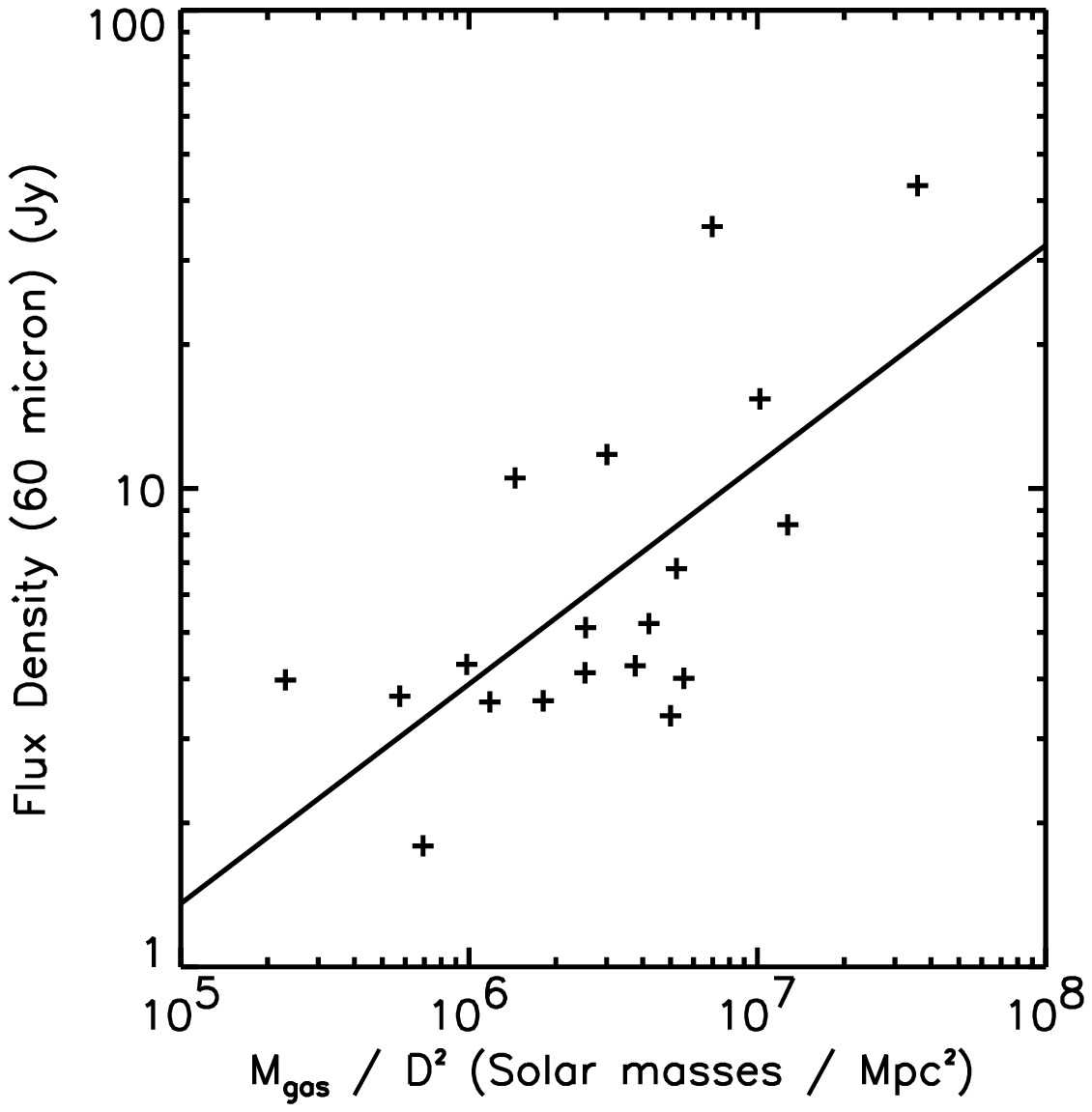,width=3.5in}
\caption{Plot of the 60~$\mu$m flux densities versus molecular gas mass for the
disks of galaxies in both the ISO and FCRAO samples.}
\label{f_disk60sch}
\end{figure}

Figures~\ref{f_disk12sch} and~\ref{f_disk60sch} show the normalized gas
masses versus 12 and 60~$\mu$m flux densities for the disk regions between
the 45 and 135~arcsec diameters of galaxies.  The trends are not as well
defined as for the nuclear regions, particularly at 12~$\mu$m.  The scatter,
particularly at low gas masses, may represent the failure of gas densities
to exceed the threshold necessary for star formation as discussed by
\citet{k89}.

In summary, these comparisons demonstrate an excellent Schmidt law relating
molecular gas and either star formation rates or far-infrared flux densities.
The trend is seen globally as well as in separate nuclear and disk regions,
with occasional deviations from the trend found in unusual objects or in
objects where gas masses fail to exceed the threshold for star formation
in their disks.

\section{Summary}

Using the ISO Atlas data for an optically-selected, magnitude-limited sample
of spiral galaxies, combined with normalization by K-band flux densities, we
have examined the evidence for star formation in three regions of these
galaxies: integrated over the entire 135~arcsec of the galaxies (global),
the central 15~arcsec (nuclear), and in the disk itself excluding the
nuclear region. We have found strong trends in star formation activity along
the Hubble sequence in all three regions with statistical significance of
$\sim$~3-6~$\sigma$. We also show that the distribution of star formation
regions is more spatially-extended in late-type galaxies than early-type
galaxies.

These results confirm most previous optical and ultraviolet investigations of
star formation along the Hubble sequence, but do not confirm the findings of an
equally large number of such studies using infrared (mostly IRAS) data, which
have found no correlations with Hubble type.  We suggest that the superior
angular  resolution in the ISO Atlas data, combined with use of K-band flux
densities for normalization, has resolved this long-standing discrepancy
between optical and infrared studies.

We argue that the differences in star formation activity along the Hubble
sequence may be understood primarily in terms of the differences in the
molecular gas density and distribution.  We confirm that star formation is
well-correlated with molecular gas mass. We suggest that the strength
of the star formation activity increases as the gas density increases along
the Hubble sequence. Similarly, we suggest that the spatial extent of star
formation activity increases as the spatial extent of the molecular gas
increases.

We found no enhancement of nuclear star formation in barred galaxies
compared to unbarred galaxies, in contrast with many previous results.
This may be because the effect is seen chiefly in early-type galaxies and
we have few early-type SA and SAB galaxies in the  sample.  We do find that
star  formation seems to be inhibited in LINERs and transition objects compared
to H~{\small II}  galaxies, suggesting that LINER emission is not powered by
recent starburst activity  (at least in nearby galaxies).

  From the far-infrared fluxes we estimate star formation rates of
1.4~M$_\sun$~yr$^{-1}$, which is in agreement with previous studies of similar
spiral galaxy samples.  However, we find gas consumption times of
6~$\times$~10$^{-8}$~yr, which is about a factor of 5 lower than most
previous studies.  We do find excellent correlation between molecular gas
densities and recent star formation activity in accord with the Schmidt Law.

\acknowledgements

The ISOPHOT data presented in this paper were reduced using PIA, which is a
joint development by the ESA Astrophysics Division and the ISOPHOT
Consortium led by MPIA, with the collaboration of the Infrared Processing and
Analysis Center (IPAC). Contributing ISOPHOT Consortium institutes are DIAS,
RAL, AIP, MPIK, and MPIA.

The ISOCAM data presented in this paper were analysed using `CIA', a joint
development by the ESA Astrophysics Division and the ISOCAM Consortium. The
ISOCAM Consortium is led by the ISOCAM PI, C. Cesarsky.

This research has been supported by NASA grants NAG 5-3370 and JPL 961566.

\clearpage

\clearpage

\begin{deluxetable}{ccc}
\tablecolumns{3}
\tablewidth{0pc}
\tablecaption{Ratio of Mid- to Far-Infrared Fluxes within $135\arcsec$ 
	Apertures for Different Hubble Types\label{t_mirfirstat}}
\tablehead{
\colhead{Type} &
\colhead{No. Galaxies} &
\colhead{Mean log($\frac{F_{MIR}}{F_{FIR}}$)}
}
\startdata
E - S0/a &	6 &	-1.05 $\pm$ 0.05 \\
Sa - Sab &	7  &	-0.94 $\pm$ 0.02 \\
Sb - Sbc &	17 &	-0.94 $\pm$ 0.03 \\
Sc - Scd &	21 &	-0.94 $\pm$ 0.03 \\
Sd - Im  &	6  &	-1.24 $\pm$ 0.14 \\
All      &	57 &	-0.98 $\pm$ 0.02 \\
\enddata
\end{deluxetable}

\begin{deluxetable}{ccccccc}
\tablecolumns{7}
\tablewidth{0pc}
\tablecaption{Star Formation Indicators for Different
	Hubble Types\label{t_sfrstat}}
\tablehead{
\colhead{Type} &
\multicolumn{2}{c}{Global Star Formation} &
\multicolumn{2}{c}{Nuclear Star Formation} &
\multicolumn{2}{c}{Disk Star Formation} \\
\colhead{} &
\colhead{No. Galaxies} &
\colhead{Mean log($\frac{F_{FIR}}{F_K}$)} &
\colhead{No. Galaxies} &
\colhead{Mean log($\frac{F_{MIR}}{F_K}$)} &
\colhead{No. Galaxies} &
\colhead{Mean log($\frac{F_{MIR}}{F_K}$)}
}
\startdata
E - S0/a &	11 &	-0.03 $\pm$ 0.10 &
         12 &  -0.92 $\pm$ 0.11 &	      6 &	-0.47 $\pm$ 0.16 \\
Sa - Sab &	7  &	0.21 $\pm$ 0.09 &
         8 &   -0.69 $\pm$ 0.10 &	      5  &	-0.06 $\pm$ 0.14 \\
Sb - Sbc &	15 &	0.69 $\pm$ 0.09 &
         17 &  -0.33 $\pm$ 0.13 &	      17 &	 0.58 $\pm$ 0.09 \\
Sc - Scd &	23 &	0.67 $\pm$ 0.05 &
         26 &  -0.30 $\pm$ 0.07 &	      21 &	 0.85 $\pm$ 0.06 \\
Sd - Im  &	7  &	0.66 $\pm$ 0.11 &
         7  &  -0.13 $\pm$ 0.14 &	      6  &	 0.87 $\pm$ 0.12 \\
All      &	63 &	0.50 $\pm$ 0.05 &
         70 &  -0.44 $\pm$ 0.06 &	      55 &	 0.54 $\pm$ 0.07 \\
\enddata
\end{deluxetable}

\clearpage

\begin{deluxetable}{cccc}
\tablecolumns{4}
\tablewidth{0pc}
\tablecaption{Results from Applying the K-S Test to Star Formation Indicators
\label{t_sfrnonpar}}
\tablehead{
\colhead{Data Sets Used} &
\multicolumn{3}{c}{Probability \tablenotemark{a}} \\
\colhead{} &
\colhead{Global log($\frac{F_{FIR}}{F_K}$)} &
\colhead{Nuclear log($\frac{F_{MIR}}{F_K}$)} &
\colhead{Disk log($\frac{F_{MIR}}{F_K}$)}}
\startdata
Sa - Sab  and  Sb - Sbc &	0.027 &		0.063 &		0.034 \\
Sa - Sab  and  Sc - Scd &	0.0010 &	0.021 &		0.0014 \\
Sb - Sbc  and  Sc - Scd &	0.33 &		0.52  &		0.083 \\
\enddata
\tablenotetext{a}{The probability of two data sets coming from the same
distribution, as determined by applying the K-S test.}
\end{deluxetable}

\begin{deluxetable}{ccc}
\tablecolumns{3}
\tablewidth{0pc}
\tablecaption{Mid-Infrared Spatial Extent Indicators for Different
	Hubble Types\label{t_mirrat}}
\tablehead{
\colhead{Type} &   \colhead{No. Galaxies} &
\colhead{Mean log($\frac{F_{MIR}(15\arcsec)}{F_{MIR}(135\arcsec)}$)}}
\startdata
E - S0/a &	4  &	-0.49 $\pm$ 0.08 \\
Sa - Sab &	5  &	-0.69 $\pm$ 0.07 \\
Sb - Sbc &	15 &	-0.99 $\pm$ 0.10 \\
Sc - Scd &	15 &	-1.21 $\pm$ 0.06 \\
Sd - Im  &	5  &	-1.11 $\pm$ 0.14 \\
All      &	44 &	-1.00 $\pm$ 0.05 \\
\enddata
\end{deluxetable}

\clearpage

\begin{deluxetable}{cc}
\tablecolumns{2}
\tablewidth{0pc}
\tablecaption{Results from Applying the K-S Test to Mid-Infrared Spatial
	Extent Indicators \label{t_mirratnonpar}}
\tablehead{
\colhead{Data Sets Used} &   \colhead{Probability \tablenotemark{a}}}
\startdata
Sa - Sab  and  Sb - Sbc &	0.23   \\
Sa - Sab  and  Sc - Scd &	0.011  \\
Sb - Sbc  and  Sc - Scd &	0.095   \\
\enddata
\tablenotetext{a}{The probability of two data sets coming from the same
distribution, as determined by applying the K-S test.}
\end{deluxetable}

\begin{deluxetable}{ccc}
\tablecolumns{3}
\tablewidth{0pc}
\tablecaption{Mid-Infrared Nuclear Star Formation Indicators for
	SA, SAB, and SB Galaxies\label{t_mirnucsfr_bar}}
\tablehead{
\colhead{Type} &   \colhead{No. Galaxies} &
\colhead{Mean log($\frac{F_{MIR}}{F_K}$)}}
\startdata
SAa - SAab &	2  &	-0.81 $\pm$ 0.38 \\
SABa - SABab &	1  &	-0.77 \\
SBa - SBab &	5  &	-0.63 $\pm$ 0.12 \\
SAb - SAbc &	6  &	-0.18 $\pm$ 0.19 \\
SABb - SABbc &	7  &	-0.46 $\pm$ 0.24 \\
SBb - SBbc &	4  &	-0.31 $\pm$ 0.28 \\
SAc - SAcd &	11  &	-0.47 $\pm$ 0.09 \\
SABc - SABcd &	9  &	-0.25 $\pm$ 0.12 \\
SBc - SBcd &	6  &	-0.08 $\pm$ 0.14 \\
\enddata
\end{deluxetable}

\clearpage

\begin{deluxetable}{cc}
\tablecolumns{2}
\tablewidth{0pc}
\tablecaption{Results from Applying the K-S Test to Mid-Infrared Nuclear
Star Formation Indicators for SA and SB Galaxies \label{t_mirnucnonpar_bar}}
\tablehead{
\colhead{Data Sets Used} &   \colhead{Probability \tablenotemark{a}}}
\startdata
SAa - SAab  and  SBa - SBab &	0.96   \\
SAb - SAbc  and  SBb - SBbc &	0.98   \\
SAc - SAcd  and  SBc - SBcd &	0.16   \\
\enddata
\tablenotetext{a}{The probability of two data sets coming from the same
distribution, as determined by applying the K-S test.}
\end{deluxetable}

\begin{deluxetable}{ccc}
\tablecolumns{3}
\tablewidth{0pc}
\tablecaption{Mid-Infrared Nuclear Star Formation Indicators for Galaxies with
	Different Nuclear Activity Types\label{t_mirnucstat_agn}}
\tablehead{
\colhead{Type} &   \colhead{No. Galaxies} &
\colhead{Mean log($\frac{F_{MIR}}{F_K}$)}}
\startdata
H~{\small II} Galaxies &	30 &	-0.26 $\pm$ 0.06 \\
Seyfert                &	3  &	-0.34 $\pm$ 0.25 \\
LINER                  &	8 &	-0.88 $\pm$ 0.11 \\
Transition Objects     &	10 &	-0.83 $\pm$ 0.12 \\
\enddata
\end{deluxetable}

\begin{deluxetable}{cc}
\tablecolumns{2}
\tablewidth{0pc}
\tablecaption{Results from the K-S Test to Mid-Infrared Nuclear
	Star Formation Indicators for Galaxies with Different Nuclear
	Activity Types\label{t_mirnucnonpar_agn}}
\tablehead{
\colhead{Data Sets Used} &   \colhead{Probability \tablenotemark{a}}}
\startdata
H~{\small II} Galaxies  and  Seyferts &			0.999   \\
H~{\small II} Galaxies  and  LINERs &			0.0023  \\
H~{\small II} Galaxies  and  Transition Objects &	0.013   \\
Seyferts                and  LINERs &			0.21	\\
LINERs                  and  Transition Objects &	0.71    \\
\enddata
\tablenotetext{a}{The probability of two data sets coming from the same
distribution, as determined by applying the K-S test.}
\end{deluxetable}

\end{document}